\documentclass[letterpaper,aps,prl,twocolumn,groupedaddress]{revtex4-1}

\usepackage{braket}
\usepackage{graphicx}
\usepackage{float}
\usepackage{amsmath}

\begin{document}

\newcommand{\Buzek}{Bu\ifmmode \check{z}\else \v{z}\fi{}ek }
\newcommand{\Buzeketal}{Bu\ifmmode \check{z}\else \v{z}\fi{}ek \textit{et al.} }
\newcommand{\Andreetal}{Andr\'{e} \textit{et al.}}

\title{Numerical test of few-qubit clock protocols}

\author{Till Rosenband}
\email{trosen@boulder.nist.gov}
\affiliation{National Institute of Standards and Technology, 325 Broadway, Boulder, CO 80305}

\date{\today}

\begin{abstract}
The stability of several clock protocols based on 2 to 20 
entangled atoms is evaluated numerically by a simulation that includes the effect of 
decoherence due to classical oscillator noise.  In this context the squeezed states 
discussed by Andr\'{e}, S\o rensen and Lukin [PRL 92, 239801 (2004)] offer reduced 
instability compared to clocks based on Ramsey's protocol with unentangled atoms.  
When more than 15 atoms are simulated, the protocol of Bu\v{z}ek, Derka and Massar 
[PRL 82, 2207 (1999)] has lower instability.  A large-scale numerical search for optimal 
clock protocols with two to eight qubits yields improved clock stability compared to 
Ramsey spectroscopy, and for two to three qubits performance matches the analytical protocols.  
In the simulations, a laser local oscillator decoheres due to flicker-frequency (1/f) 
noise.  The oscillator frequency is repeatedly corrected, based on projective 
measurements of the qubits, which are assumed not to decohere with one another.
\end{abstract}

\maketitle

\section{Introduction}
Atomic clocks are intrinsically quantum measurement devices, 
and it is an open question to what degree quantum many-body 
states can improve clock operation.  Squeezed states were 
first discussed in the context of optical interferometers 
with improved resolution~\cite{JOSAB1987SqueezedLight}.  
Subsequently, spin-squeezed input states~\cite{Kitagawa1993Squeezed} 
were considered for improved frequency resolution in atomic 
clocks~\cite{Wineland1994Squeezed,JJB1996Entanglement}.  
Further studies simultaneously optimized the initial quantum state 
with the clock's measurement basis~\cite{Buzek1999Optimal} 
to achieve frequency resolution that scales as the Heisenberg limit. 

In atomic clocks, the highest accuracies are currently reached in the optical frequency domain with pairs of trapped 
ions~\cite{CWC2010AlAl} where the atom-number is difficult to increase without loss of 
accuracy.  Experiments of similar construction have demonstrated arbitrary unitary 
transformations of ion-qubit pairs~\cite{Hanneke2010Programmable}.  Therefore, ``quantum gain'', where 
improved performance is extracted from a small number of entangled atoms in clocks, may 
have practical significance.

This study focuses on the projection-noise-limited frequency stability of passive 
atomic clocks~\footnote{In passive atomic clocks a classical oscillator is 
frequency-stabilized to an atomic resonance.  In active atomic clocks such as masers, 
the clock signal is produced directly by the atoms.  Active optical clocks may have a 
different type of noise floor~\cite{Meiser2009Superradiance}.}, where the classical 
oscillator is the only source of decoherence, and the atomic qubits are assumed not to 
decohere with one another.  This situation has been been addressed for the case of 
squeezed states with large qubit numbers~\cite{Lukin2004Stability}, and for general 
quantum states and measurement bases, also in the limit of large qubit 
numbers~\cite{Buzek1999Optimal}.  Experimental trapped-atom optical clocks that are 
based on resonances of metastable excitations share this decoherence mechanism when 
inter-atom decoherence from interactions, spontaneous emission, and background-field 
fluctuations can be neglected.  In optical atomic clocks the oscillator frequency is 
derived from laser-stabilization cavities that have an intrinsic thermal noise 
floor~\cite{Numata2004ThermalNoise} whose power-spectrum of frequency-fluctuations 
scales as $1/f$.  This fundamental thermal-noise floor serves as the model for 
oscillator decoherence.  Note that models of decoherence and oscillator noise are 
essential ingredients for studies of optical-clock stability, where the uncertainty 
of the atom-oscillator phase difference is typically of order one radian.  In contrast, 
the atom-oscillator phase difference in many microwave atomic clocks is of order one 
milliradian, and oscillator noise plays a different role.

\section{Clock model}
The simulated clocks are generalizations of Ramsey's clock protocol~\cite{Ramsey1956}.  Each clock contains $N$ qubits whose states are the ground state $\ket{0}=\bigl(\begin{smallmatrix} 1\\ 0 \end{smallmatrix} \bigr)$ and the excited state $\ket{1}=\bigl(\begin{smallmatrix} 0\\ 1 \end{smallmatrix} \bigr)$.  Ramsey's protocol, as considered here, consists of repeated application of the following steps where $\pi/2$ rotations are assumed to be infinitesimally short:

\begin{enumerate}
\item Prepare initial state. All qubits are placed in the state $\psi_1=\bigl(\begin{smallmatrix} 1\\ -\mathrm{i} \end{smallmatrix} \bigr)/\sqrt{2}$ which is the state $\bigl(\begin{smallmatrix} 1\\ 0 \end{smallmatrix} \bigr)$  after rotation by $\pi/2$ about the Bloch-sphere $x$-axis.  
\item Free evolution for a period $T$ where a phase difference of $\phi$ accumulates between the oscillator and the qubits. $\psi_2=\bigl(\begin{smallmatrix} 1&0\\ 0&e^{-\mathrm{i}\phi} \end{smallmatrix} \bigr)\psi_1$
\item Measure final state by rotating the qubits by $\pi/2$ about the Bloch-sphere $y$-axis, and counting the number of excited-state qubits.  This corresponds to measuring $\psi_2$ in the basis $a_1=\bigl(\begin{smallmatrix} 1\\ -1 \end{smallmatrix} \bigr)/\sqrt{2}$, 
$a_2=\bigl(\begin{smallmatrix} 1\\ 1 \end{smallmatrix} \bigr)/\sqrt{2}$.
\item Adjust the oscillator frequency by an amount that depends on the measurement outcome in step 3.
\item Add a random variable to the oscillator frequency to model its $1/f$ noise floor.  The oscillator frequency has a probe cycle to probe cycle variance of $2~\mathrm{Hz}^2$, independent of $T$, corresponding to a flat Allan deviation of $1$~Hz.  This noise level is chosen for convenience, and is of similar magnitude to the experimental oscillator noise in optical clocks.
\end{enumerate}

In this description the Bloch-sphere rotation directions are defined by the oscillator phase.  Thus, when the oscillator accumulates a phase error during the free evolution period, this is modeled as the phase $\phi$ that is applied differentially to the two states of each qubit in step 2.    

The above sequence can be understood as a measurement of the atom-oscillator frequency difference in steps 1 to 3, followed by a correction of the oscillator frequency.  Atomic projection noise in step 3 limits the measurement stability to~\cite{WMI1993ProjectionNoise} 
\begin{equation}
\sigma_f(\tau)=\frac{1}{2\pi\sqrt{N T\tau}},
\end{equation}
where $\sigma_f(\tau)$ is the standard deviation of the clock frequency after it has been 
averaged over the period $\tau$, with respect to the true frequency of the $N$ atoms.  
In order to minimize  $\sigma_f$, one should maximize the free evolution period $T$.  
However, when $T$ is too large, it is possible for frequency errors of $\pm2\pi/T$ to 
accumulate undetected, because the atomic signal is periodic.  Such occurrences, called 
``fringe hops,'' limit the duration of $T$.  Note that clock protocols with variable $T$ 
may avoid fringe hops and allow for improved stability.  Beam clocks naturally avoid this difficulty due to their thermal velocity spread~\cite{Ramsey1990Nobel}.

Ramsey's protocol can be generalized in two ways.  The first is the use of arbitrary multi-qubit states for $\psi_1$.  When these states reduce the phase measurement uncertainty in the limit of small $T$, the states are considered spin-squeezed~\cite{Kitagawa1993Squeezed,Wineland1994Squeezed}.  Andr\'{e} \emph{et al.}~\cite{Lukin2004Stability} suggest the states $\psi_1=\mathcal{N}(\kappa)\sum_{m=-N/2}^{N/2} (-1)^m e^{-(m/\kappa)^2}\ket{N,m+N/2}$, where $\kappa$ parameterizes the degree of spin-squeezing, $\mathcal{N}(\kappa)$ provides normalization, and the states $\ket{N,m}$ are the fully-symmetrized states of $N$ qubits containing $m$ excitations.  For example, $\ket{4,3}=(\ket{0111} + \ket{1011} + \ket{1101} + \ket{1110})/2$.  These initial states improve the stability of simulated clocks (see RESULTS).  The second generalization consists of the use of other measurement bases.  Only initial states and measurement bases in the symmetric subspace spanned by the states $\ket{N,m}$ are considered~\cite{Buzek1999Optimal}.  Such protocols consist of repeated application of these steps:

\begin{enumerate}
\item Prepare initial state $\psi_1$.
\item Free evolution for a time-period $T$. $\ket{N,m}\rightarrow e^{-\mathrm{i}m\phi}\ket{N,m}$
\item Measure final state by projecting into a measurement basis $\{\ket{a_j}\}$.
\item Adjust the oscillator frequency by an amount that depends on which $\ket{a_j}$ was measured in step 3.
\end{enumerate}

Oscillator noise is simulated as before.  This protocol could be further generalized to include the possibility of partial measurements, ancilla qubits, and frequency corrections that depend also on the measurement outcomes from prior cycles.  However, such extensions are not considered here.  Furthermore, the free-evolution period $T$ is fixed for each protocol instance.

\Buzeketal have optimized analytically the initial state $\psi_1$ and basis $\{\ket{a_j}\}$ for phase measurements in the limit of large $N$.  The authors find  
\begin{equation}
\ket{\psi_1}=\sum_{m=0}^{N} \sqrt{\frac{2}{N+1}}\sin{\frac{\pi(m+1/2)}{N+1}} \ket{N,m}
\end{equation}
and
\begin{equation}
\ket{a_j} = \frac{1}{\sqrt{N+1}}\sum_{m=0}^{N}e^{i m \phi(j)}\ket{N,m},
\end{equation}
where $\phi(j)=\frac{2\pi j}{N+1}$.
Simulated clocks based on this protocol achieve 
Heisenberg-limited scaling (see RESULTS).  The phase-shifted 
basis states where $\phi(j)=\frac{2\pi (j+1/2)}{N+1}$ are also considered, 
because they offer reduced clock instability when $N$ is odd.
 
The above optimization is for a uniform distribution of oscillator 
phase errors on the interval $[-\pi, \pi)$, while in experimental 
clocks the phase error has a distribution that is peaked at $\phi=0$ 
and drops near zero as $\phi$ approaches $\pm\pi$.  The authors also assume that phase errors can be considered modulo $2\pi$ but when the free-evolution period $T$ is optimized in optical clocks, phase errors beyond $2\pi$ must also be taken into account.
In this work the frequency corrections associated with each basis state $\ket{a_j}$
are numerically optimized, to account for the 
non-uniform distribution at the correction stage.  Other studies 
explicitly optimize phase estimation protocols as a function of 
the prior distribution of phases~\cite{Demkowicz2011BeyondQFisher, Mullan2011Clocks}, and this approach is likely to result in more stable few-qubit clock protocols.

\section{Numerical search}
The generalized protocol described above can be parameterized by an array of real numbers.  For each protocol, the expected instability can then be calculated in a Monte Carlo simulation.  In the present study, a numerical optimizer adjusts the protocol parameters to find the best performance.  This is a difficult numerical problem, because the dimensionality grows quickly with qubit number $N$, and numerical optimizers are not well suited to optimize the results of Monte Carlo simulations, which contain noise from the randomization process.  Performance estimates based on Markov chains would avoid the problem of randomization noise, but because the 1/f noise process is non-stationary, a very large state space may be needed for accurate estimates of long-term stability.  Nevertheless, the present numerical search yields protocols whose performance exceeds that of Ramsey's protocol for two to eight qubits.

Minimization of the search-space dimensionality is critical.  
As noted above, only initial states and measurement bases in the 
symmetric subspace (spanned by the $\ket{N,m}$ states) are considered.  
For the numerical search, $2N+1$ real numbers (reals) parameterize the 
initial state and $N^2+N$ reals parameterize the measurement basis. 
Because both unitary operations and measurement bases can be written 
as orthonormal matrices, their parameterization is nearly identical.  
Efficient parameterization of unitary operations is described by Tilma 
and Sudarshan~\cite{Tilma2002SUN}.  For the present calculation, 
extraneous phase degrees of freedom have been removed from the basis states.  
In addition, $N+1$ reals parameterize the frequency corrections, and one 
real parameterizes the free-evolution period $T$.  In total, $N^2+4N+3$ 
reals parameterize an $N$-qubit clock.  It is believed that this exceeds the minimal 
parameterization by one real.

Clock performance is measured as the long-term instability.  That is, if the clock runs for many interrogation cycles, how close is the average oscillator-frequency to that of the atomic qubits?  The Monte Carlo simulator propagates the clock through $10^5$ cycles, and calculates the variance of 100-cycle frequency-averages.  Additional steps are taken to ensure that this variance reflects the long-term clock instability. The oscillator noise is pre-computed to have a $1/f$ power spectrum of frequency noise~\cite{Numata2004ThermalNoise,Lennon2000Noise} with an Allan deviation~\cite{Riley2008Stability} of 1~Hz.

It should be noted that for a fixed free-evolution period $T$, the frequency of all clocks with finite $N$ diverges as a random walk, because undetectable $2\pi$ phase jumps (fringe hops) cannot be avoided entirely.  Nevertheless the probability of fringe hops can be made small for large but finite numbers of clock cycles.  This regime of a large, but finite number of clock cycles describes both real clocks, which do not run forever, and the Monte Carlo simulations in this work. 

The search was performed via Nelder-Mead optimization~\cite{Galassi2006GSL} of randomized protocols that meet a performance threshold.  All $N^2+4N+3$ parameters were randomly varied for the general search, so that all possible initial states and measurement bases were within the search space.  When initial tests yielded good stability, the protocol was run through the optimizer for further refinement.  For known protocols, only the frequency corrections and free-evolution period $T$ were varied, as well as the squeezing parameter $\kappa$, where applicable.  For all protocols, optimal frequency corrections were initialized by assuming a prior Gaussian distribution of frequency errors, and computing the mean frequency associated with each possible measurement outcome $\ket{a_j}$.  To these estimates random offsets were added before testing the protocol.  In the case of known protocols, certain symmetries are evident, and these symmetries were also enforced for the frequency corrections.
\begin{figure}
\includegraphics[width=1.0\linewidth]{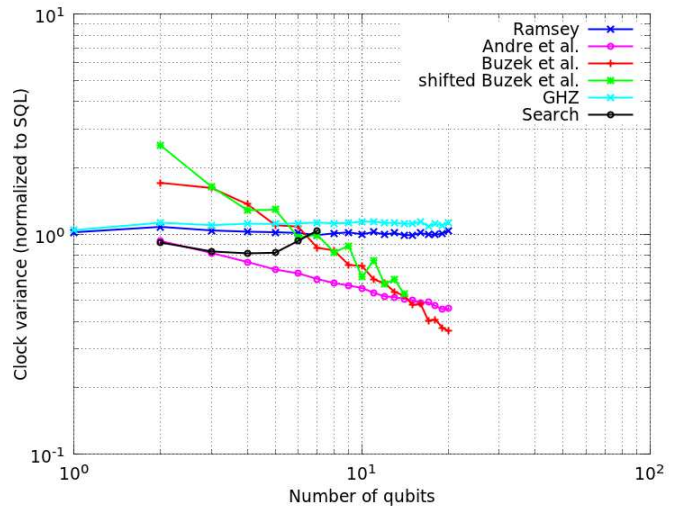}
\caption{(color online) Long-term statistical variance of entangled clocks that contain different numbers of qubits, compared to the standard quantum limit (SQL).  The most stable clocks found by the large-scale search are shown as black points.  Each point is based on several hours of runtime on NIST’s computing cluster, where typically 2000 processor cores were utilized in parallel.  Also shown is the simulated performance of analytically optimized clock protocols.  Approximately 15 qubits are required to improve upon the SQL by a factor of two.}\label{figPerf}
\end{figure}

\section{Results}
\begin{figure*}[ht!]
\includegraphics[width=0.9\textwidth]{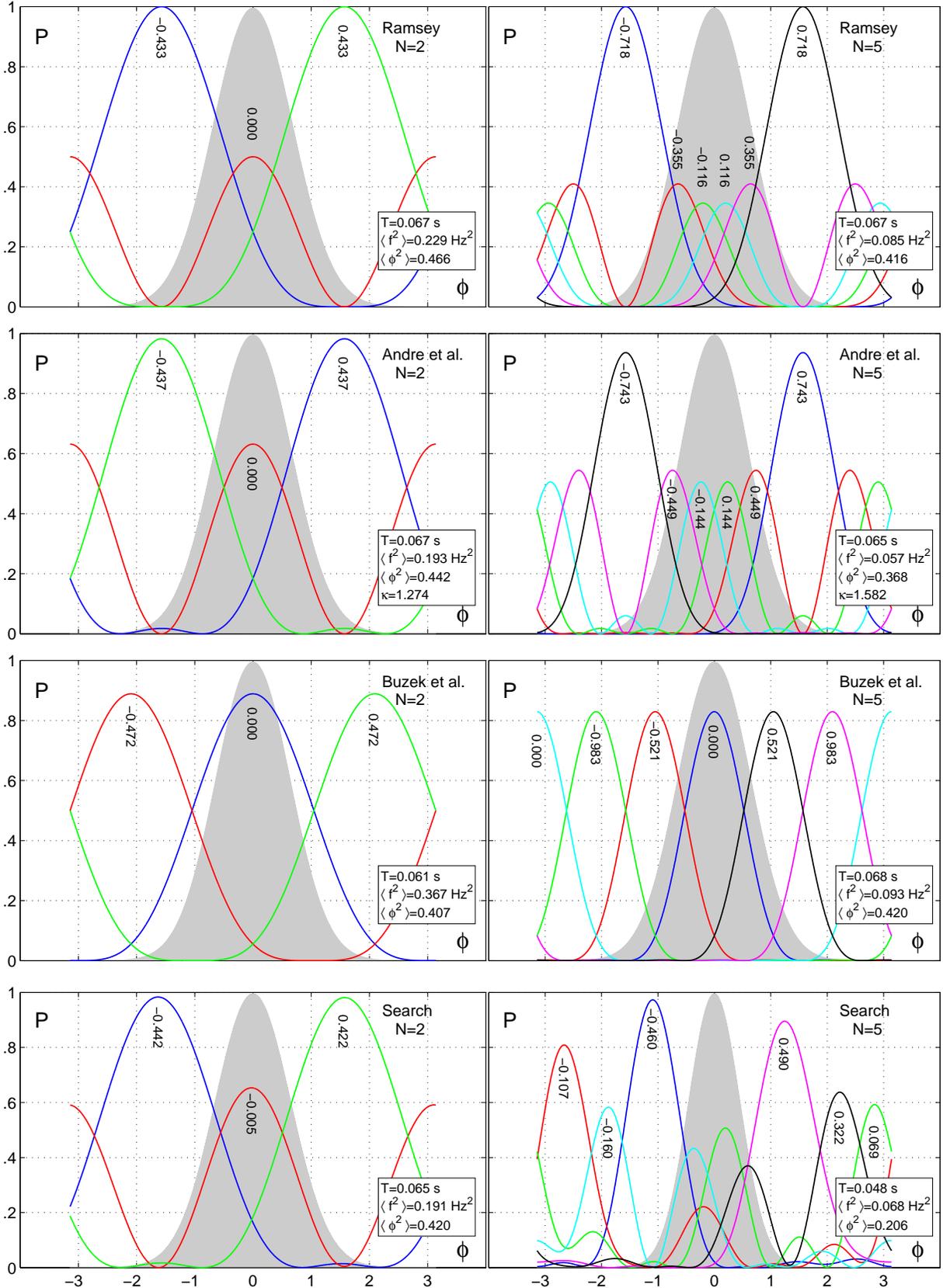}
\caption{(color) Probability (P) of measuring each basis state as a 
function of the atom-oscillator phase difference ($\phi$).  Shown are the 
various protocols for two and five atoms.  Each differently 
colored curve corresponds to a basis state that $\psi_1$ is projected onto 
after free evolution.  Vertical text near the curves' peaks indicates the 
optimized phase estimate ($\phi_{Est}$).  In the simulations, the frequency 
corrections are $\phi_{Est}/(2\pi T)$.  Shaded in the background is the 
Gaussian distribution whose variance $\langle\phi^2\rangle$ represents the 
atom-oscillator phase differences that occur in the simulation.  Also listed 
is the optimized probe period $T$, squeezing parameter $\kappa$ where 
applicable, and long-term frequency variance of the clock extrapolated to 
1 second. For long-term averages of $n$ seconds, the variance is 
$\langle f^2\rangle/n$.}\label{figProt}
\end{figure*}

Numerical simulations of the clock protocols considered here are summarized 
in Figure~\ref{figPerf}.  Ramsey's protocol defines the standard quantum limit (SQL), 
and it is evident that entangled states of two or more qubits can reduce 
clock instability, although the GHZ states~\cite{JJB1996Entanglement} yield no gain for the noise model considered here, as has been noted previously~\cite{DJW1998bible, TR2008FSM}.  The spin-squeezed states suggested by Andr\'{e} 
\emph{et al.} yield the best performance for 3 to 15 qubits, and improve 
upon the SQL variance by a factor of $N^{-1/3}$.  For more qubits, the 
protocol of \Buzeketal further reduces clock variance, because this 
protocol scales as $N^{-1}$.  The numerical search was run in many 
parellel threads to find protocols that surpass these analytical 
protocols, but for two qubits, the clock-variance is only reduced by 1~\%, 
within the margin of error for this calculation.  An example of the search result for two qubits is
\begin{eqnarray*}
U&=\left( \begin{array}{ccc} 
-0.486- 0.039\mathrm{i} &  0.708- 0.132\mathrm{i} &  0.335+ 0.363\mathrm{i}\\ 
 0.470+ 0.106\mathrm{i} &  0.687- 0.082\mathrm{i} & -0.364- 0.395\mathrm{i}\\ 
-0.570+ 0.454\mathrm{i} & -0.043+ 0.000\mathrm{i} & -0.684+ 0.000\mathrm{i} \end{array} \right)\\
\psi_1&=\left(\begin{array}{c}-0.572+ 0.000\mathrm{i}\\-0.220- 0.580\mathrm{i}\\
-0.458- 0.282\mathrm{i}\end{array}\right)
\end{eqnarray*}
where Fig.~\ref{figProt}~(bottom-left) includes the corrections and probe period, and probability amplitudes for the different possible measurement outcomes can be written as 
$U e^{-\mathrm{i}\hat{H}T/\hbar} \psi_1$.
\begin{figure}
\includegraphics[width=1.0\linewidth]{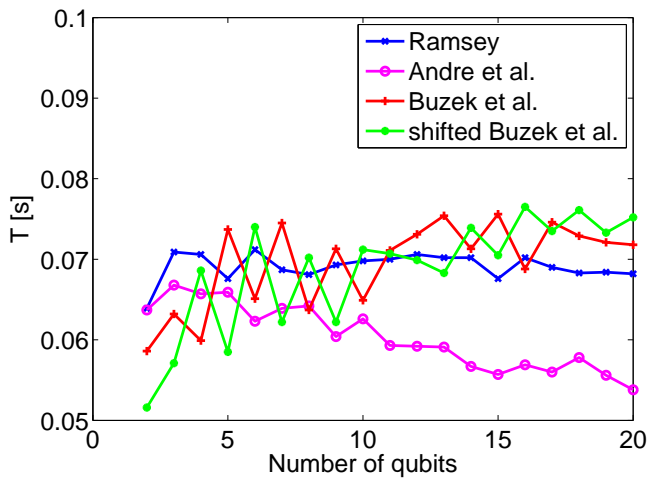}
\caption{(color) Numerically optimized free-evolution period $T$ for some of clock protocols considered here, when the oscillator noise has an Allan deviation of 1~Hz.}\label{figT}
\end{figure}
Although the analytical protocols are within the search space, 
their performance is not reached by the general search program for $N>3$, 
due to the size of the problem. The optimized free-evolution period $T$ for different 
protocols is shown in Fig.~\ref{figT}.
The behavior of the different types of clocks is illustrated in Fig.~\ref{figProt}, 
where it can be seen that the protocol of \Buzeketal gains frequency 
resolution as $1/N$, because each basis state coprreponds to a range of 
phases that shrinks as $1/N$.  The squeezed-state protocols are similar to 
Ramsey's protocol, but gain frequency resolution near $\phi=0$ at the 
expense of decreased resolution near $\phi=\pm\pi/2$.  For $N=2$ the 
``Search'' protocol is very similar to that of \Andreetal

\section{Conclusion}
For accurate ion clocks, further experimental improvements are required to achieve 
full quantum control of two clock qubits, where the available ``quantum gain'' 
appears to be 15~\% to 20~\%.  It is likely that gains of similar magnitude 
can be derived from easier to implement classical improvements, where the 
free-evolution period is varied to prevent fringe hops.  Such classical 
protocols will define a new standard quantum limit with which to compare 
variable probe time, entanglement-based, protocols.  It remains an open 
question how much quantum gain is possible in variable probe-time clocks.  
Recent theoretical work on efficient quantum-phase estimation~\cite{Demkowicz2011BeyondQFisher, Mullan2011Clocks} 
may improve upon the protocols considered here~\cite{Buzek1999Optimal,Lukin2004Stability}, 
especially for three or more qubits.

\section*{Acknowledgments}
Helpful discussions with S.~Lloyd, E.~Knill, D.~L.~Rosenband, and 
D.~J.~Wineland are gratefully acknowledged.  This work was supported by 
the DARPA QuASaR program, ONR, and AFOSR.  Contribution of NIST, not subject to U.S. copyright.

\bibliography{../TR}

\begin{thebibliography}{10}%
\makeatletter
\providecommand \@ifxundefined [1]{%
 \ifx #1\undefined \expandafter \@firstoftwo
 \else \expandafter \@secondoftwo
\fi
}%
\providecommand \@ifnum [1]{%
 \ifnum #1\expandafter \@firstoftwo
 \else \expandafter \@secondoftwo
\fi
}%
\providecommand \enquote [1]{``#1''}%
\providecommand \bibnamefont  [1]{#1}%
\providecommand \bibfnamefont [1]{#1}%
\providecommand \citenamefont [1]{#1}%
\providecommand\href[0]{\@sanitize\@href}%
\providecommand\@href[1]{\endgroup\@@startlink{#1}\endgroup\@@href}%
\providecommand\@@href[1]{#1\@@endlink}%
\providecommand \@sanitize [0]{\begingroup\catcode`\&12\catcode`\#12\relax}%
\@ifxundefined \pdfoutput {\@firstoftwo}{%
 \@ifnum{\z@=\pdfoutput}{\@firstoftwo}{\@secondoftwo}%
}{%
 \providecommand\@@startlink[1]{\leavevmode\special{html:<a href="#1">}}%
 \providecommand\@@endlink[0]{\special{html:</a>}}%
}{%
 \providecommand\@@startlink[1]{%
  \leavevmode
  \pdfstartlink
   attr{/Border[0 0 1 ]/H/I/C[0 1 1]}%
   user{/Subtype/Link/A<</Type/Action/S/URI/URI(#1)>>}%
  \relax
 }%
 \providecommand\@@endlink[0]{\pdfendlink}%
}%
\providecommand \url  [0]{\begingroup\@sanitize \@url }%
\providecommand \@url [1]{\endgroup\@href {#1}{\urlprefix}}%
\providecommand \urlprefix [0]{URL }%
\providecommand \Eprint[0]{\href }%
\@ifxundefined \urlstyle {%
  \providecommand \doi [1]{doi:\discretionary{}{}{}#1}%
}{%
  \providecommand \doi [0]{doi:\discretionary{}{}{}\begingroup
  \urlstyle{rm}\Url }%
}%
\providecommand \doibase [0]{http://dx.doi.org/}%
\providecommand \Doi[1]{\href{\doibase#1}}%
\providecommand \bibAnnote [3]{%
  \BibitemShut{#1}%
  \begin{quotation}\noindent
    \textsc{Key:}\ #2\\\textsc{Annotation:}\ #3%
  \end{quotation}%
}%
\providecommand \bibAnnoteFile [2]{%
  \IfFileExists{#2}{\bibAnnote {#1} {#2} {\input{#2}}}{}%
}%
\providecommand \typeout [0]{\immediate \write \m@ne }%
\providecommand \selectlanguage [0]{\@gobble}%
\providecommand \bibinfo [0]{\@secondoftwo}%
\providecommand \bibfield [0]{\@secondoftwo}%
\providecommand \translation [1]{[#1]}%
\providecommand \BibitemOpen[0]{}%
\providecommand \bibitemStop [0]{}%
\providecommand \bibitemNoStop [0]{.\EOS\space}%
\providecommand \EOS [0]{\spacefactor3000\relax}%
\providecommand \BibitemShut [1]{\csname bibitem#1\endcsname}%
\bibitem{JOSAB1987SqueezedLight}%
  \BibitemOpen
  \bibinfo {note} {JOSA B \textbf{4}, 1450 (1987), \textit{Special issue on
  squeezed light} (ed. H.~J.~Kimble and D.~F.~Walls)}%
  \bibAnnoteFile{NoStop}{JOSAB1987SqueezedLight}%
\bibitem{Kitagawa1993Squeezed}%
  \BibitemOpen
  \bibfield{author}{%
  \bibinfo {author} {\bibfnamefont{M.}~\bibnamefont{Kitagawa}}\ and\ \bibinfo
  {author} {\bibfnamefont{M.}~\bibnamefont{Ueda}},\ }%
  \bibfield{journal}{%
  \Doi{10.1103/PhysRevA.47.5138}{\bibinfo {journal} {Phys. Rev. A}}\ }%
  \textbf{\bibinfo {volume} {47}},\ \bibinfo {pages} {5138} (\bibinfo {year}
  {1993})%
  \bibAnnoteFile{NoStop}{Kitagawa1993Squeezed}%
\bibitem{Wineland1994Squeezed}%
  \BibitemOpen
  \bibfield{author}{%
  \bibinfo {author} {\bibfnamefont{D.~J.}\ \bibnamefont{Wineland}}, \bibinfo
  {author} {\bibfnamefont{J.~J.}\ \bibnamefont{Bollinger}}, \bibinfo {author}
  {\bibfnamefont{W.~M.}\ \bibnamefont{Itano}},\ and\ \bibinfo {author}
  {\bibfnamefont{D.~J.}\ \bibnamefont{Heinzen}},\ }%
  \bibfield{journal}{%
  \Doi{10.1103/PhysRevA.50.67}{\bibinfo {journal} {Phys. Rev. A}}\ }%
  \textbf{\bibinfo {volume} {50}},\ \bibinfo {pages} {67} (\bibinfo {year}
  {1994})%
  \bibAnnoteFile{NoStop}{Wineland1994Squeezed}%
\bibitem{JJB1996Entanglement}%
  \BibitemOpen
  \bibfield{author}{%
  \bibinfo {author} {\bibfnamefont{J.~J.}\ \bibnamefont{Bollinger}}, \bibinfo
  {author} {\bibfnamefont{W.~M.}\ \bibnamefont{Itano}}, \bibinfo {author}
  {\bibfnamefont{D.~J.}\ \bibnamefont{Wineland}},\ and\ \bibinfo {author}
  {\bibfnamefont{D.~J.}\ \bibnamefont{Heinzen}},\ }%
  \bibfield{journal}{%
  \Doi{10.1103/PhysRevA.54.R4649}{\bibinfo {journal} {Phys. Rev. A}}\ }%
  \textbf{\bibinfo {volume} {54}},\ \bibinfo {pages} {R4649} (\bibinfo {year}
  {1996})%
  \bibAnnoteFile{NoStop}{JJB1996Entanglement}%
\bibitem{Buzek1999Optimal}%
  \BibitemOpen
  \bibfield{author}{%
  \bibinfo {author} {\bibfnamefont{V.}~\bibnamefont{Bu\ifmmode~\check{z}\else
  \v{z}\fi{}ek}}, \bibinfo {author} {\bibfnamefont{R.}~\bibnamefont{Derka}},\
  and\ \bibinfo {author} {\bibfnamefont{S.}~\bibnamefont{Massar}},\ }%
  \bibfield{journal}{%
  \Doi{10.1103/PhysRevLett.82.2207}{\bibinfo {journal} {Phys. Rev. Lett.}}\ }%
  \textbf{\bibinfo {volume} {82}},\ \bibinfo {pages} {2207} (\bibinfo {year}
  {1999})%
  \bibAnnoteFile{NoStop}{Buzek1999Optimal}%
\bibitem{CWC2010AlAl}%
  \BibitemOpen
  \bibfield{author}{%
  \bibinfo {author} {\bibfnamefont{C.~W.}\ \bibnamefont{Chou}}, \bibinfo
  {author} {\bibfnamefont{D.~B.}\ \bibnamefont{Hume}}, \bibinfo {author}
  {\bibfnamefont{J.~C.~J.}\ \bibnamefont{Koelemeij}}, \bibinfo {author}
  {\bibfnamefont{D.~J.}\ \bibnamefont{Wineland}},\ and\ \bibinfo {author}
  {\bibfnamefont{T.}~\bibnamefont{Rosenband}},\ }%
  \bibfield{journal}{%
  \Doi{10.1103/PhysRevLett.104.070802}{\bibinfo {journal} {Phys. Rev. Lett.}}\
  }%
  \textbf{\bibinfo {volume} {104}},\ \bibinfo {pages} {070802} (\bibinfo {year}
  {2010})%
  \bibAnnoteFile{NoStop}{CWC2010AlAl}%
\bibitem{Hanneke2010Programmable}%
  \BibitemOpen
  \bibfield{author}{%
  \bibinfo {author} {\bibfnamefont{D.}~\bibnamefont{{Hanneke}}}, \bibinfo
  {author} {\bibfnamefont{J.~P.}\ \bibnamefont{{Home}}}, \bibinfo {author}
  {\bibfnamefont{J.~D.}\ \bibnamefont{{Jost}}}, \bibinfo {author}
  {\bibfnamefont{J.~M.}\ \bibnamefont{{Amini}}}, \bibinfo {author}
  {\bibfnamefont{D.}~\bibnamefont{{Leibfried}}},\ and\ \bibinfo {author}
  {\bibfnamefont{D.~J.}\ \bibnamefont{{Wineland}}},\ }%
  \bibfield{journal}{%
  \Doi{10.1038/nphys1453}{\bibinfo {journal} {Nature Physics}}\ }%
  \textbf{\bibinfo {volume} {6}},\ \bibinfo {pages} {13} (\bibinfo {year}
  {2010})%
  \bibAnnoteFile{NoStop}{Hanneke2010Programmable}%
\bibitem{Note1}%
  \BibitemOpen
  \bibinfo {note} {In passive atomic clocks a classical oscillator is
  frequency-stabilized to an atomic resonance. In active atomic clocks such as
  masers, the clock signal is produced directly by the atoms. Active optical
  clocks may have a different type of noise floor~\cite
  {Meiser2009Superradiance}.}%
  \bibAnnoteFile{Stop}{Note1}%
\bibitem{Lukin2004Stability}%
  \BibitemOpen
  \bibfield{author}{%
  \bibinfo {author} {\bibfnamefont{A.}~\bibnamefont{Andr\'e}}, \bibinfo
  {author} {\bibfnamefont{A.~S.}\ \bibnamefont{S\o{}rensen}},\ and\ \bibinfo
  {author} {\bibfnamefont{M.~D.}\ \bibnamefont{Lukin}},\ }%
  \bibfield{journal}{%
  \Doi{10.1103/PhysRevLett.92.230801}{\bibinfo {journal} {Phys. Rev. Lett.}}\
  }%
  \textbf{\bibinfo {volume} {92}},\ \bibinfo {pages} {230801} (\bibinfo {year}
  {2004})%
  \bibAnnoteFile{NoStop}{Lukin2004Stability}%
\bibitem{Numata2004ThermalNoise}%
  \BibitemOpen
  \bibfield{author}{%
  \bibinfo {author} {\bibfnamefont{K.}~\bibnamefont{Numata}}, \bibinfo {author}
  {\bibfnamefont{A.}~\bibnamefont{Kemery}},\ and\ \bibinfo {author}
  {\bibfnamefont{J.}~\bibnamefont{Camp}},\ }%
  \bibfield{journal}{%
  \Doi{10.1103/PhysRevLett.93.250602}{\bibinfo {journal} {Phys.\ Rev.\ Lett.}}\
  }%
  \textbf{\bibinfo {volume} {93}},\ \bibinfo {pages} {250602} (\bibinfo {year}
  {2004})%
  \bibAnnoteFile{NoStop}{Numata2004ThermalNoise}%
\bibitem{Ramsey1956}%
  \BibitemOpen
  \bibfield{author}{%
  \bibinfo {author} {\bibfnamefont{N.~F.}\ \bibnamefont{{Ramsey}}},\ }%
  \emph{\bibinfo {title} {Molecular Beams}}\ (\bibinfo {publisher} {Oxford
  University Press},\ \bibinfo {year} {1956})%
  \bibAnnoteFile{NoStop}{Ramsey1956}%
\bibitem{WMI1993ProjectionNoise}%
  \BibitemOpen
  \bibfield{author}{%
  \bibinfo {author} {\bibfnamefont{W.~M.}\ \bibnamefont{Itano}}, \bibinfo
  {author} {\bibfnamefont{J.~C.}\ \bibnamefont{Bergquist}}, \bibinfo {author}
  {\bibfnamefont{J.~J.}\ \bibnamefont{Bollinger}}, \bibinfo {author}
  {\bibfnamefont{J.~M.}\ \bibnamefont{Gilligan}}, \bibinfo {author}
  {\bibfnamefont{D.~J.}\ \bibnamefont{Heinzen}}, \bibinfo {author}
  {\bibfnamefont{F.~L.}\ \bibnamefont{Moore}}, \bibinfo {author}
  {\bibfnamefont{M.~G.}\ \bibnamefont{Raizen}},\ and\ \bibinfo {author}
  {\bibfnamefont{D.~J.}\ \bibnamefont{Wineland}},\ }%
  \bibfield{journal}{%
  \bibinfo {journal} {Phys.\ Rev.\ A}\ }%
  \textbf{\bibinfo {volume} {47}},\ \bibinfo {pages} {3554} (\bibinfo {year}
  {1993})%
  \bibAnnoteFile{NoStop}{WMI1993ProjectionNoise}%
\bibitem{Ramsey1990Nobel}%
  \BibitemOpen
  \bibfield{author}{%
  \bibinfo {author} {\bibfnamefont{N.~F.}\ \bibnamefont{Ramsey}},\ }%
  \bibfield{journal}{%
  \Doi{10.1103/RevModPhys.62.541}{\bibinfo {journal} {Rev. Mod. Phys.}}\ }%
  \textbf{\bibinfo {volume} {62}},\ \bibinfo {pages} {541} (\bibinfo {month}
  {Jul}\ \bibinfo {year} {1990})%
  \bibAnnoteFile{NoStop}{Ramsey1990Nobel}%
\bibitem{Demkowicz2011BeyondQFisher}%
  \BibitemOpen
  \bibfield{author}{%
  \bibinfo {author} {\bibfnamefont{R.}~\bibnamefont{{Demkowicz-Dobrzanski}}},\
  }%
  \bibinfo {note} {arXiv:1102.0786 (2011)}%
  \bibAnnoteFile{NoStop}{Demkowicz2011BeyondQFisher}%
\bibitem{Mullan2011Clocks}%
  \BibitemOpen
  \bibfield{author}{%
  \bibinfo {author} {\bibfnamefont{M.}~\bibnamefont{{Mullan}}}\ and\ \bibinfo
  {author} {\bibfnamefont{E.}~\bibnamefont{{Knill}}},\ }%
  \bibinfo {note} {arXiv:1107.5347 (2011)}%
  \bibAnnoteFile{NoStop}{Mullan2011Clocks}%
\bibitem{Tilma2002SUN}%
  \BibitemOpen
  \bibfield{author}{%
  \bibinfo {author} {\bibfnamefont{T.}~\bibnamefont{{Tilma}}}\ and\ \bibinfo
  {author} {\bibfnamefont{E.~C.~G.}\ \bibnamefont{{Sudarshan}}},\ }%
  \bibfield{journal}{%
  \Doi{10.1088/0305-4470/35/48/316}{\bibinfo {journal} {J. Phys. A}}\ }%
  \textbf{\bibinfo {volume} {35}},\ \bibinfo {pages} {10467} (\bibinfo {year}
  {2002})%
  \bibAnnoteFile{NoStop}{Tilma2002SUN}%
\bibitem{Lennon2000Noise}%
  \BibitemOpen
  \bibfield{author}{%
  \bibinfo {author} {\bibfnamefont{J.~L.}\ \bibnamefont{Lennon}},\ }%
  \bibfield{journal}{%
  \bibinfo {journal} {Ecography}\ }%
  \textbf{\bibinfo {volume} {23}},\ \bibinfo {pages} {101} (\bibinfo {year}
  {2000})%
  \bibAnnoteFile{NoStop}{Lennon2000Noise}%
\bibitem{Riley2008Stability}%
  \BibitemOpen
  \bibfield{author}{%
  \bibinfo {author} {\bibfnamefont{W.~J.}\ \bibnamefont{Riley}},\ }%
  \emph{\bibinfo {title} {Handbook of Frequency Stability Analysis}}\ (\bibinfo
  {year} {2008})\ \bibinfo {note} {, NIST Spec. Pub. 1065}%
  \bibAnnoteFile{NoStop}{Riley2008Stability}%
\bibitem{Galassi2006GSL}%
  \BibitemOpen
  \bibfield{author}{%
  \bibinfo {author} {\bibfnamefont{M.}~\bibnamefont{Galassi}} \emph{et~al.},\
  }%
  \emph{\bibinfo {title} {GNU Scientific Library Reference Manual - 3rd Ed.}}\
  (\bibinfo {year} {2009})\ ISBN \bibinfo {isbn} {0-9541617-3-4}%
  \bibAnnoteFile{NoStop}{Galassi2006GSL}%
\bibitem{DJW1998bible}%
  \BibitemOpen
  \bibfield{author}{%
  \bibinfo {author} {\bibfnamefont{D.~J.}\ \bibnamefont{Wineland}}, \bibinfo
  {author} {\bibfnamefont{C.}~\bibnamefont{Monroe}}, \bibinfo {author}
  {\bibfnamefont{W.~M.}\ \bibnamefont{Itano}}, \bibinfo {author}
  {\bibfnamefont{D.}~\bibnamefont{Leibfried}}, \bibinfo {author}
  {\bibfnamefont{B.~E.}\ \bibnamefont{King}},\ and\ \bibinfo {author}
  {\bibfnamefont{D.~M.}\ \bibnamefont{Meekhof}},\ }%
  \bibfield{journal}{%
  \bibinfo {journal} {J.\ Res.\ NIST}\ }%
  \textbf{\bibinfo {volume} {103}},\ \bibinfo {pages} {259} (\bibinfo {year}
  {1998})%
  \bibAnnoteFile{NoStop}{DJW1998bible}%
\bibitem{TR2008FSM}%
  \BibitemOpen
  \bibfield{author}{%
  \bibinfo {author} {\bibfnamefont{T.}~\bibnamefont{Rosenband}} \emph{et~al.},\
  }%
  in\ \emph{\bibinfo {booktitle} {Proceedings of the 7th Symposium on Frequency
  Standards and Metrology}}\ (\bibinfo {publisher} {World Scientific},\
  \bibinfo {year} {2008})%
  \bibAnnoteFile{NoStop}{TR2008FSM}%
\bibitem{Meiser2009Superradiance}%
  \BibitemOpen
  \bibfield{author}{%
  \bibinfo {author} {\bibfnamefont{D.}~\bibnamefont{Meiser}}, \bibinfo {author}
  {\bibfnamefont{J.}~\bibnamefont{Ye}}, \bibinfo {author}
  {\bibfnamefont{D.~R.}\ \bibnamefont{Carlson}},\ and\ \bibinfo {author}
  {\bibfnamefont{M.~J.}\ \bibnamefont{Holland}},\ }%
  \bibfield{journal}{%
  \Doi{10.1103/PhysRevLett.102.163601}{\bibinfo {journal} {Phys. Rev. Lett.}}\
  }%
  \textbf{\bibinfo {volume} {102}},\ \bibinfo {pages} {163601} (\bibinfo {year}
  {2009})%
  \bibAnnoteFile{NoStop}{Meiser2009Superradiance}%
\end{thebibliography}%

\end{document}